**No winners: Performance of lung cancer prediction models depends on screening-detected, incidental, and biopsied pulmonary nodule use cases.**

Thomas Z. Li[a,b], BS, Kaiwen Xu[c], BS, MSc, Aravind Krishnan[d], BE, Riqiang Gao[e], PhD, Michael N. Kammer[f], PhD, Sanja Antic[f], MD, David Xiao[g], MD, Michael Knight[f], MD, Yency Martinez[f], MD, Rafael Paez[f], MD, MSCI, Robert J. Lentz[f], MD, Stephen Deppen[g], PhD, Eric L. Grogan[g], MD, MPH, Thomas A. Lasko[c,h], MD, PhD, Kim L. Sandler[i], MD, Fabien Maldonado[f], MD, MSc, Bennett A. Landman[b,c,d,i], PhD

[a]Medical Scientist Training Program, Vanderbilt University, Nashville, 37235, TN, USA

[b]Biomedical Engineering, Vanderbilt University, Nashville, 37235, TN, USA

[c]Computer Science, Vanderbilt University, Nashville, 37235, TN, USA

[d]Electrical and Computer Engineering, Vanderbilt University, Nashville, 37235, TN, USA

[e]Digital Technology and Innovation, Siemens Healthineers, Princeton NJ 08540, USA

[f]Division of Allergy, Pulmonary and Critical Care Medicine, Department of Medicine, Vanderbilt University Medical Center, Nashville, 37232, TN, USA

[g]Department of Thoracic Surgery, Vanderbilt University Medical Center, Nashville, 37232, TN, USA

[h]Biomedical Informatics, Vanderbilt University Medical Center, Nashville, 37232, TN, USA

[i]Radiology and Radiological Sciences, Vanderbilt University Medical Center, Nashville, 37232, TN, USA

**Abstract**

**Purpose:** Statistical models for predicting lung cancer have the potential to facilitate earlier diagnosis of malignancy and avoid invasive workup of benign disease. Many models have been published, but comparative studies of their utility in different clinical settings in which patients would arguably most benefit are scarce.

**Materials and Methods:** This study retrospectively evaluated promising predictive models for lung cancer prediction in three clinical settings: lung cancer screening with low-dose computed tomography, incidentally detected pulmonary nodules, and nodules deemed suspicious enough to warrant a biopsy. We leveraged 9 cohorts (n=898, 896, 882, 219, 364, 117, 131, 115, 373) from multiple institutions to assess the area under the receiver operating characteristic curve (AUC) of validated models including logistic regressions on clinical variables and radiologist nodule characterizations, artificial intelligence on chest CTs, longitudinal imaging AI, and multi-modal approaches. We implemented each model from their published literature, re-training the models if necessary, and curated each cohort from primary data sources.

**Results:** We observed that model performance varied greatly across clinical use cases. No single predictive model emerged as a clear winner across all cohorts, but certain models excelled in specific clinical contexts. Single timepoint chest CT AI performed well in lung screening, but struggled to generalize to other clinical settings. Longitudinal imaging and multimodal models demonstrated comparatively promising performance on incidentally-detected nodules. However, when applied to nodules that underwent biopsy, all models underperformed.

**Conclusion:** These results underscore the strengths and limitations of 8 validated predictive models and highlight promising directions towards personalized, noninvasive lung cancer diagnosis.

# 1. Introduction

Every year, an estimated 1.57 million Americans have at least one pulmonary nodule detected either on routine chest computed tomographic (CT) in incidental fashion or during lung cancer screening [1]. While biopsy of the nodule remains the gold standard diagnostic test for malignancy, it is done with an invasive procedure associated with morbidity, mortality, healthcare costs and anxiety for patients [2,3]. With 95% of indeterminate pulmonary nodules (IPNs) being benign [4], clinical guidelines [1,5–7] recommend to risk-stratify nodules before resorting to invasive diagnostics and surgical intervention. Statistical models for predicting lung cancer have the potential to improve this risk-stratification, aiding earlier diagnosis of malignancy as well as reducing morbidity, costs, and anxiety associated with workup of benign disease.

Validated predictive models developed to stratify pulmonary nodules consist of (1) clinical prediction models, (2) cross-sectional or longitudinal AI models, and (3) multimodal approaches. We consider a predictive model validated if it has demonstrated competitive discriminatory performance (above 0.75 AUC) on a separate test cohort. The Brock [8] and Mayo [9] models are two of the most used models in clinical practice and recommended by clinical guidelines. They are well-validated logistic regressions and are based on readily available variables such as demographics [10], smoking history, and radiologists' assessment of nodule features. However, they require radiologists to first detect and characterize the nodule, a step that can be subject to inter-reader variability [11–13].

Recent research has validated several AI models for cancer prediction. These operate directly on the voxels of the image, negating the need for radiologists to first describe nodule morphology or measure sizes. One of the early AI successes in lung cancer prediction was Liao et al [14]. Their two-step approach involved first detecting suspicious lesions in the lung field from a single chest CT and then computing malignancy risk from the proposed regions of interest (ROIs). Recent work by Mikhael et al.[15] publicly released Sybil, a predictive model that extracts global chest features along with regional attention features to predict lung cancer risk up to 6-years.

Previous work has also leveraged AI on longitudinal imaging. Gao et al.[16] and Li et al.[17] extended Liao et al. to leverage consecutive chest CTs and the time interval between scans. In another longitudinal imaging approach, Ardila et al.[18], demonstrated impressive area under the receiver operating characteristic (AUC) performance by including global chest features outside of local ROIs, but their model was not released publicly. Recently, efforts that leveraged data from multiple modalities have shown [19,20], with limited validation, that the combination of clinical variables and imaging AI can improve performance over single modality approaches.

The plethora of research activity is promising, but several concerns arise when considering the clinical utility of predictive models in lung cancer diagnostics. First, the comparative advantage of AI models versus commonly used linear models has not been quantitative characterized in settings where they would arguably most benefit. Second, almost all of the AI models are, to some extent, trained on lung screening scans from the National Lung Screening Trial (NLST) [21], which raises the question of whether they generalize across institutions and to patients with incidentally-detected nodules and metastases to the lung from other sites. Third, these models predict different outcomes. Some assess the risk of developing lung cancer over a multi-year period whereas others estimate the probability that an observed pulmonary nodule is malignant. A comparative analysis of these models using a standardized outcome (e.g. 2-year diagnosis of lung cancer) can inform "off-label" usage of models, but has not been done. Finally, there is an urgent need to risk-stratify intermediate-risk nodules and reduce the number of biopsies on nodules that appear indeterminate but turn out to be benign. To this end, there has not been a systematic analysis of how existing models perform in this setting.

In this study we address these questions through an evaluation of 8 validated predictive models on cohorts with screening-detected nodules, incidentally-detected nodules collected in both retrospective and prospective fashion, and nodules that underwent a bronchoscopic biopsy. These different settings are where we envision a well-designed predictive model will have a tangible impact on patient care. We implemented each model from their published code repositories and curated each cohort from their primary available source. We suggest that this analysis can inform appropriate usage of models in important clinical settings, highlight situations when these models fail, and offer strategies that lead to highly generalized performance.

## 2. Materials and Methods

### 2.1 Cohorts

The data included in this study were sourced from the National Lung Screening Trail (NLST) and our medical center A This study also included data from a multi-institute consortium which includes the VA associated with medical center A, medical center B, Detection of Early lung Cancer Among Military Personnel (DECAMP), and medical center C. We derived 10 named cohorts from these sites using different inclusion criteria (Table 1) and provide a summary of their characteristics (Table 2). We obtained CT scans, demographics, questionnaire data from CT arm of the NLST upon request with a data use agreement (https://cdas.cancer.gov/learn/nlst/images/). Data from LS-A was acquired under our home institutional review board (IRB) supervision #181279. LI-A and BRONCH were acquired under IRB supervision #140274. Consortium-A, was acquired under IRB supervision #030763 and #000616. Consortium-B, Consortium-C, and DECAMP were acquired via academic collaborations under different grants.

### 2.2 Image Preprocessing

We ensured that all CT scans used in this study passed certain image quality standards. Specifically, we excluded scans with and severe imaging artifacts, scans with slice thickness greater than or equal to 5 mm, and scans without the full lung field in the field of view [22]. Patient health information was removed using the MIRC Anonymizer [23].

**2.3 Predictive Models**

We selected an array of different approaches to lung cancer prediction (Table 2) that included models designed to estimate lung cancer risk (i.e. Sybil) as well as models designed to predict the malignancy probability of a pulmonary nodule. We included Brock and Mayo because they are among the most cited, validated, and used in clinical practice. We studied a several AI models incorporate a range of approaches that would allow us to examine the efficacy of three strategies: single timepoint chest CT, longitudinal chest CT, and multi-modality input.

We split subjects with confirmed follow-up in the NLST into NLST-dev and NLST-test. NLST-test contains the subjects in the Ardila et al. test set that had confirmed follow-up and these scans remained unseen until evaluation. NLST-dev was used to retrain, from random weights, several of the models using a standardized 2-year lung cancer outcome. The purpose of retraining was to (1) ensure that the models were blinded to NLST-test and (2) standardize the predicted outcome across each model. Specifically, the years between initial observation of the subject and the outcome, or *year-to-outcome*, was not standardized across the evaluated models (Table 3). Models developed using a shorter year-to-outcome have an easier task than models developed using a longer year-to-outcome because predictions are more difficult the further out they are. In this way, differences in year-to-outcomes can confound model comparisons. For longitudinal imaging models, the outcome was whether the subject was diagnosed with lung cancer within two years of the subject's latest scan. The logistic regression models were not retrained and were evaluated as published since they were already blinded to NLST-test and we did not have their original development dataset which is needed to control for year-to-outcome. Sybil was also evaluated as published because the model was already blinded to NLST-test and its prediction includes a 2-year outcome. Lastly, DeepLungIPN was originally trained using a cross-validation of Consortium-A, Consortium-DECAMP, and Consortium-C. This model was evaluated as published since the model includes a blood biomarker that was only collected in the MCL cohorts.

Model implementation and training remained faithful to the published literature. Details on the site of development and training distribution are reported in Supp. Table 3. Apart from removing scans that did not meet our image quality standards, we did not add or remove any pre- or post-processing steps included in the models' pipeline.

**2.4 Evaluation**

Evaluation include all the named cohorts except NLST-dev. A model-cohort evaluation was not feasible when a substantial portion of the input data was missing (Supp. Table 1). We did not conduct an evaluation when more than 10% of a cohort's subjects were missing an input variable. When a variable was missing in less than 10% of subjects, we conducted an evaluation using imputation based on a multivariate regression of the other available variables. The effect of imputation on model performance is evaluated in Supp. Table 2. We did not evaluate longitudinal imaging models (DLSTM and TDViT) on cohorts where longitudinal imaging was unavailable. When evaluating DeepLungIPN on the consortium cohorts, we report the out-of-fold cross validation results.

We used area under the receiver operating curve (AUC) to measure model performance on classifying lung cancer cases and benign controls. For each model-cohort evaluation, we used bootstrapping procedure to estimate the model's performance on the cohort's true population. The procedure drew 1000 samples of the same size with replacement from the original cohort. Each model's AUC was calculated for each sample and we reported the mean AUC and 95% confidence interval (CI) over all bootstrapped samples. A two-sided Wilcoxon-signed rank test evaluated significance of differences in mean AUC between models within a single cohort. We did not test statistical differences across cohorts because subjects were not paired across cohorts.

Model performance on non-small cell lung cancer (NSCLC) vs. small cell lung cancer (SCLC) cases were compared in NLST-test-nodules and Consortium-A. The mean AUC and 95% CI of each model was computed using the same bootstrap procedure drawn from the pool of benign subjects and subjects with either SCLC or NSCLC. An unpaired t-test evaluated whether a model's discrimination of malignant versus benign was significantly different with these two subtypes.

An analysis of calibration before and after confidence correction was conducted. In each model-cohort evaluation, 10-fold cross validation was used to fit isotonic regressions on the training set of each fold. Calibration was then evaluated on the validation set of each fold (Supp. Fig. 3-6).

## 4. Results

Table 4 reports the mean AUC and 95% CI for each feasible model-cohort evaluation. Comparing the results row-wise reveals that almost all predictive models exhibited noticeable differences in performance across cohorts (Figure 1A). The performance gaps were the largest between cohorts from different sites and different clinical settings (i.e. Brock on NLST-test-nodules: 0.789 [0.782, 0.796] vs. Brock on Consortium-DECAMP: 0.624 [0.621, 0.627]). The performance gap remained large between cohorts from different sites but the same clinical setting (i.e. Sybil on NLST-test: 0.881 [0.877, 0.885] vs. on LS-A: 0.779 [0.768, 0.789]). In contrast, the performance gap between different cohorts from the same site but different clinical setting was generally smaller (i.e. Liao et al. on LS-A: 0.723 [0.711, 0.734] vs. on Consortium-A: 0.662 [0.660, 0.664]).

Comparing the relative performances between multiple models across cohorts highlights a few findings:

- Longitudinal or multimodal AI were top performers across all cohorts with incidental nodules (Figure 2).
- Single chest CT AI performed well in lung cancer screening cohorts (i.e. Sybil on NLST-test: 0.881 [0.877, 0.885]).
- Models evaluated on the BRONCH cohort, representing a nodules that are suspicious enough to warrant a biopsy, performed poorly with mean AUCs ranging from 0.50 to 0.62 (Figure 3C).
- Single chest CT AI were generally competitive with linear models while longitudinal and multi-modal AI significantly outperformed linear models in every cohort. Results for Consortium-B represent this well, with Sybil (0.889 [0.884, 0.895]) performing close to Brock (0.873 [0.871, 0.875]), and DeepLungIPN (0.943 [0.941, 0.944]) outstripping the performance of both.

- Longitudinal and multimodal models showed better worst case performances in comparison to the other approaches (Figure 1B). Ranking the results within each cohort, we define a model's worst case as its lowest ranked performance across all cohorts except BRONCH. The worst-case performances of DLSTM (on NLST-test-nodules: 0.727 [0.721, 0.731]), TDViT (on NLST-test-nodules: 0.797 [0.793, 0.802]), DeepLungScreening (on NLST-test-nodules: 0.774 [0.769, 0.778]), and DeepLungIPN (on Consortium-DECAMP: 0.767 [0.765, 0.769]) were all moderate in terms of absolute AUC. In contrast, the worst-case performance of Mayo, Brock, Liao et al., and Sybil were low in terms of absolute AUC.

- AI models discriminated NSCLC cases from benign better than SCLC cases in the lung screening setting. In both lung screening and incidental-detected nodules, longitudinal models demonstrated better performance with NSCLC cases compared with SCLC cases (Table 5).

## 5. Discussion

This study presents a systematic evaluation of statistical models and their performance three clinically relevant settings: patients undergoing lung cancer screening, patients with incidentally detected nodules, and patients with nodules recommended for biopsy. Cohorts were selected from a several institutions and time frames, which offered insight into site-generalizability of each model. The selected models covered four broad categories of statistical approaches: logistic regression on expert-collected clinical variables, single chest CT AI, longitudinal chest CT AI, and multimodal AI.

The most prominent result was perhaps that there was no clear winner among the models evaluated (Figure 2). The performance of each model varied with site and clinical setting, which reflects a moderate degree of generalization failure that is often observed in both open sourced and commercial predictive models across many medical domains [24]. Those interested in using predictive models in lung cancer should be aware that these models, despite previous reports of successful external validation, most reliably achieve their expected performance when they are used in the same clinical context and site as they were developed in [25]. Those involved in model deployment should consider fine-tuning models with a cohort that matches the site, clinical setting, and year-to-outcome in which the model will be used. Steps should be taken during model development to mitigate a failure to generalize when site and setting are unmatched with techniques such as image harmonization [26], fine-tuning [27], and potentially directly modeling the site-specific effects [24]. These results motivate further investigation into the site- and context-specific factors that are driving a variance in performance and how they can be harmonized.

This study reveals the importance of interpreting a model's performance relative to the performance of other models on the same cohort. Doing so revealed several findings that were sustained across cohorts.

Single chest CT AI (Liao et al. and Sybil) performed on par with linear models that included nodule variables (Mayo and Brock). In its current form, single chest CT AI is well suited for identifying individuals at risk for lung cancer who can benefit from starting or having more frequent lung imaging (Figure 3A).

Longitudinal and multimodal models demonstrated comparatively favorable performance on incidentally-detected nodules (Figure 3B). In contrast to other models, longitudinal and multimodal AI also appeared to be more robust across cohorts, as seen from their worst-case performances (Figure 1B).

Given that nodules in the BRONCH cohort were inherently difficult to diagnose, the poor performance on this cohort was unsurprising (Figure 3C). Due to missing data, we were not able to evaluate longitudinal and multimodal AI on BRONCH. A predictive model that is highly specific for lung cancer in this setting has the potential to prevent invasive management of benign nodules. Therefore, evaluation of longitudinal and multimodal AI on a retrospective cohort of biopsied nodules is a high priority area for future investigation.

Longitudinal imaging models performed better on NSCLC than SCLC. One explanation for this is that NSCLC is, on average, observed more frequently as an indeterminant nodule compared to faster progressing SCLC which is often advanced stage at first observation [28]. These results warn that longitudinal imaging models may underperform on SCLC cases.

The regression calibrator improved calibration for most models evaluated on the NSLT, Consortium, and BRONCH cohorts. Calibration remained poor or became worse for models evaluated on highly imbalanced cohorts (LS-A and LI-A), which align with previous findings [29].

Within the AI approaches, leveraging additional sources complementary data appears to be an effective strategy for improving classification performance. For instance, Sybil makes use of the entire chest CT whereas Liao et al. predicts on a few regions of interests (ROIs), a technique that crops out portions of the lung field and discards the overall chest anatomy. Additionally, using longitudinal imaging when, when available, leads to performance gains in across most of the cohorts. The integration of two or more consecutive chest CTs allows the model to consider how imaging features change over time. The use of data from multiple modalities also appears to be effective. From a clinical perspective, the advantage of multimodality is expected, as imaging findings are often interpreted in the context of the patient's clinical risk factors. The improved performance of longitudinal AI and multimodal AI in this study suggest that combining the two approaches is a promising direction.

We note the following limitations of this study. Since the evaluation cohorts are a few years dated, we expect different numerical results on cohorts drawn from today's practice but a similar failure to generalize across clinical context and site. Several model-cohort evaluations were not conduced due to incomplete data. Extreme class imbalance in the models' training cohort is another confounding factor that can affect a model's sensitivity and specificity. This is concerning for the Brock model which was trained on a cohort with a cancer prevalence much smaller than those of other models. Other confounding sources include the differences in cohort size, scanner manufacturers, and scanner protocols [30,31]. Finally, the evaluation of DeepLungIPN on its training cohort is limited because the results are from cross-validation. However, it still performed well when evaluated on a true external cohort (Consortium-B).

In summary, this study presents a comparative analysis of 8 predictive models against 9 cohorts that represent clinical relevant use cases. Our results revealed a lack of generalized performance and that certain modeling strategies excelled in lung screening vs. incidentally-detected nodules, while all models fell short in a cohort with biopsied nodules. We highlight approaches in lung cancer predictive modeling that, if investigated further, have the potential to overcome these observed limitations.

## 6. Acknowledgements.

We extend our gratitude to Dr. Heidi Chen for their statistical guidance in service of this research.

**Funding**

This research was funded by the NIH through F30CA275020, 2U01CA152662, and R01CA253923-02, as well as NSF CAREER 1452485 and NSF 2040462. This study was also funded by the Vanderbilt Institute for Surgery and Engineering through T32EB021937-07, the Vanderbilt Institute for Clinical and Translational Research through UL1TR002243-06, and the Pierre Massion Directorship in Pulmonary Medicine.

## 7. References

[1] M.P. Rivera, A.C. Mehta, M.M. Wahidi, Establishing the Diagnosis of Lung Cancer: Diagnosis and Management of Lung Cancer, 3rd ed: American College of Chest Physicians Evidence-Based Clinical Practice Guidelines, Chest 143 (2013) e142S-e165S. https://doi.org/10.1378/CHEST.12-2353.

[2] T. Lokhandwala, M.A. Bittoni, R.A. Dann, A.O. D'Souza, M. Johnson, R.J. Nagy, R.B. Lanman, R.E. Merritt, D.P. Carbone, Costs of Diagnostic Assessment for Lung Cancer: A Medicare Claims Analysis, Clin Lung Cancer 18 (2017) e27–e34. https://doi.org/10.1016/j.cllc.2016.07.006.

[3] P.P. Massion, R.C. Walker, Indeterminate Pulmonary Nodules: Risk for Having or for Developing Lung Cancer?, Cancer Prevention Research 7 (2014) 1173–1178. https://doi.org/10.1158/1940-6207.CAPR-14-0364.

[4] M.K. Gould, T. Tang, I.L.A. Liu, J. Lee, C. Zheng, K.N. Danforth, A.E. Kosco, J.L. Di Fiore, D.E. Suh, Recent trends in the identification of incidental pulmonary nodules, Am J Respir Crit Care Med 192 (2015) 1208–1214. https://doi.org/10.1164/RCCM.201505-0990OC.

[5] P.J. Mazzone, L. Lam, Evaluating the Patient With a Pulmonary Nodule: A Review, JAMA 327 (2022) 264–273. https://doi.org/10.1001/JAMA.2021.24287.

[6] H. MacMahon, D.P. Naidich, J.M. Goo, K.S. Lee, A.N.C. Leung, J.R. Mayo, A.C. Mehta, Y. Ohno, C.A. Powell, M. Prokop, G.D. Rubin, C.M. Schaefer-Prokop, W.D. Travis, P.E. Van Schil, A.A. Bankier, Guidelines for management of incidental pulmonary nodules detected on CT images: From the Fleischner Society 2017, Radiology 284 (2017) 228–243. https://doi.org/10.1148/RADIOL.2017161659.

[7] F.C. Detterbeck, S.Z. Lewis, R. Diekemper, D. Addrizzo-Harris, W.M. Alberts, Executive Summary: Diagnosis and Management of Lung Cancer, 3rd ed: American College of Chest Physicians Evidence-Based Clinical Practice Guidelines, Chest 143 (2013) 7S-37S. https://doi.org/10.1378/chest.12-2377.

[8] S.J. Swensen, M.D. Silverstein, D.M. Ilstrup, C.D. Schleck, E.S. Edell, The Probability of Malignancy in Solitary Pulmonary Nodules: Application to Small Radiologically Indeterminate Nodules, Arch Intern Med 157 (1997) 849–855. https://doi.org/10.1001/ARCHINTE.1997.00440290031002.

[9] A. McWilliams, M.C. Tammemagi, J.R. Mayo, H. Roberts, G. Liu, K. Soghrati, K. Yasufuku, S. Martel, F. Laberge, M. Gingras, S. Atkar-Khattra, C.D. Berg, K. Evans, R. Finley, J. Yee, J. English, P. Nasute, J. Goffin, S. Puksa, L. Stewart, S. Tsai, M.R. Johnston, D. Manos, G. Nicholas, G.D. Goss, J.M. Seely, K. Amjadi, A. Tremblay, P. Burrowes, P. MacEachern, R. Bhatia, M.-S.

Tsao, S. Lam, Probability of Cancer in Pulmonary Nodules Detected on First Screening CT, New England Journal of Medicine 369 (2013) 910–919. https://doi.org/10.1056/NEJMOA1214726.

[10] M.C. Tammemägi, H.A. Katki, W.G. Hocking, T.R. Church, N. Caporaso, P.A. Kvale, A.K. Chaturvedi, G.A. Silvestri, T.L. Riley, J. Commins, C.D. Berg, Selection criteria for lung-cancer screening, N Engl J Med 368 (2013) 728–736. https://doi.org/10.1056/NEJMOA1211776.

[11] M.P. Revel, A. Bissery, M. Bienvenu, L. Aycard, C. Lefort, G. Frija, Are Two-dimensional CT Measurements of Small Noncalcified Pulmonary Nodules Reliable?, Radiology 231 (2004) 453–458. https://doi.org/10.1148/RADIOL.2312030167.

[12] G.R. Oxnard, B. Zhao, C.S. Sima, M.S. Ginsberg, L.P. James, R.A. Lefkowitz, P. Guo, M.G. Kris, L.H. Schwartz, G.J. Riely, Variability of lung tumor measurements on repeat computed tomography scans taken within 15 minutes, Journal of Clinical Oncology 29 (2011) 3114–3119. https://doi.org/10.1200/JCO.2010.33.7071.

[13] A. Devaraj, B. Van Ginneken, A. Nair, D. Baldwin, Use of volumetry for lung nodule management: Theory and practice1, Radiology 284 (2017) 630–644. https://doi.org/10.1148/RADIOL.2017151022.

[14] F. Liao, M. Liang, Z. Li, X. Hu, S. Song, Evaluate the Malignancy of Pulmonary Nodules Using the 3D Deep Leaky Noisy-or Network, IEEE Trans Neural Netw Learn Syst 30 (2017) 3484–3495. https://doi.org/10.1109/tnnls.2019.2892409.

[15] P.G. Mikhael, J. Wohlwend, A. Yala, L. Karstens, J. Xiang, A.K. Takigami, P.P. Bourgouin, P. Chan, S. Mrah, W. Amayri, Y.-H. Juan, C.-T. Yang, Y.-L. Wan, G. Lin, L. V. Sequist, F.J. Fintelmann, R. Barzilay, Sybil: A Validated Deep Learning Model to Predict Future Lung Cancer Risk From a Single Low-Dose Chest Computed Tomography, Journal of Clinical Oncology (2023). https://doi.org/10.1200/jco.22.01345.

[16] R. Gao, Y. Tang, K. Xu, Y. Huo, S. Bao, S.L. Antic, E.S. Epstein, S. Deppen, A.B. Paulson, K.L. Sandler, P.P. Massion, B.A. Landman, Time-distanced gates in long short-term memory networks, Med Image Anal 65 (2020) 101785. https://doi.org/10.1016/J.MEDIA.2020.101785.

[17] T.Z. Li, K. Xu, R. Gao, Y. Tang, T.A. Lasko, F. Maldonado, K.L. Sandler, B.A. Landman, Time-distance vision transformers in lung cancer diagnosis from longitudinal computed tomography, Https://Doi.Org/10.1117/12.2653911 12464 (2023) 229–238. https://doi.org/10.1117/12.2653911.

[18] D. Ardila, A.P. Kiraly, S. Bharadwaj, B. Choi, J.J. Reicher, L. Peng, D. Tse, M. Etemadi, W. Ye, G. Corrado, D.P. Naidich, S. Shetty, End-to-end lung cancer screening with three-dimensional deep learning on low-dose chest computed tomography, Nature Medicine 2019 25:6 25 (2019) 954–961. https://doi.org/10.1038/s41591-019-0447-x.

[19] R. Gao, Y. Tang, K. Xu, M. Kammer, S. Antic, al Riqiang Gao, M.N. Kammer, S.L. Antic, S. Deppen, K.L. Sandler, P.P. Massion, Y. Huo, B.A. Landman, Deep multi-path network integrating incomplete biomarker and chest CT data for evaluating lung cancer risk, Https://Doi.Org/10.1117/12.2580730 11596 (2021) 387–393. https://doi.org/10.1117/12.2580730.

[20] R. Gao, Y. Tang, M.S. Khan, K. Xu, A.B. Paulson, S. Sullivan, Y. Huo, S. Deppen, P.P. Massion, K.L. Sandler, B.A. Landman, Cancer risk estimation combining lung screening ct with clinical data elements, Radiol Artif Intell 3 (2021). https://doi.org/10.1148/RYAI.2021210032.

[21] Reduced Lung-Cancer Mortality with Low-Dose Computed Tomographic Screening, New England Journal of Medicine 365 (2011) 395–409. https://doi.org/10.1056/NEJMOA1102873.

[22] R. Gao, M.S. Khan, Y. Tang, K. Xu, S. Deppen, Y. Huo, K.L. Sandler, P.P. Massion, B.A. Landman, Technical Report: Quality Assessment Tool for Machine Learning with Clinical CT, (n.d.). https://www.vumc.org/radiology/lung (accessed June 13, 2023).

[23] The MIRC DICOM Anonymizer - MircWiki, (n.d.). https://mircwiki.rsna.org/index.php?title=The_MIRC_DICOM_Anonymizer

[24] T.A. Lasko, E. V. Strobl, W.W. Stead, Why Do Clinical Probabilistic Models Fail To Transport Between Sites?, (2023). https://arxiv.org/abs/2311.04787v1

[25] Alexey Youssef, Michael Pencina, Anshul Thakur, Tingting Zhu, David Clifton, Nigam H. Shah, External validation of AI models in health should be replaced with recurring local validation, Nature Medicine 2023 29:11 29 (2023) 2686–2687. https://doi.org/10.1038/s41591-023-02540-z.

[26] F. Hu, A.A. Chen, H. Horng, V. Bashyam, C. Davatzikos, A. Alexander-Bloch, M. Li, H. Shou, T.D. Satterthwaite, M. Yu, R.T. Shinohara, Image harmonization: A review of statistical and deep learning methods for removing batch effects and evaluation metrics for effective harmonization, Neuroimage 274 (2023) 120125. https://doi.org/10.1016/J.NEUROIMAGE.2023.120125.

[27] T.W. Li, G.C. Lee, Performance Analysis of Fine-tune Transferred Deep Learning, Proceedings of the 3rd IEEE Eurasia Conference on IOT, Communication and Engineering 2021, ECICE 2021 (2021) 315–319. https://doi.org/10.1109/ECICE52819.2021.9645649.

[28] C.M. Rudin, E. Brambilla, C. Faivre-Finn, J. Sage, Small-cell lung cancer, Nature Reviews Disease Primers 2021 7:1 7 (2021) 1–20. https://doi.org/10.1038/s41572-020-00235-0.

[29] T.M.S.F.P.Flach. Meelis Kull, Beyond sigmoids: How to obtain well-calibrated probabilities from binary classifiers with beta calibration., Electron J Stat 11 (2017) 5052–5080.

[30] Y. Li, L. Lu, M. Xiao, L. Dercle, Y. Huang, Z. Zhang, L.H. Schwartz, D. Li, B. Zhao, CT Slice Thickness and Convolution Kernel Affect Performance of a Radiomic Model for Predicting EGFR Status in Non-Small Cell Lung Cancer: A Preliminary Study, Scientific Reports 2018 8:1 8 (2018) 1–10. https://doi.org/10.1038/s41598-018-36421-0.

[31] J. Choe, S.M. Lee, K.H. Do, G. Lee, J.G. Lee, S.M. Lee, J.B. Seo, Deep Learning–based Image Conversion of CT Reconstruction Kernels Improves Radiomics Reproducibility for Pulmonary Nodules or Masses, Radiology 292 (2019) 365–373. https://doi.org/10.1148/RADIOL.2019181960.

[32] T.Z. Li, K. Xu, N.C. Chada, H. Chen, M. Knight, S. Antic, K.L. Sandler, F. Maldonado, B.A. Landman, T.A. Lasko, Curating Retrospective Multimodal and Longitudinal Data for Community Cohorts at Risk for Lung Cancer, MedRxiv (2023) 2023.11.03.23298020. https://doi.org/10.1101/2023.11.03.23298020.

[33] E. Billatos, F. Duan, E. Moses, H. Marques, I. Mahon, L. Dymond, C. Apgar, D. Aberle, G. Washko, A. Spira, Detection of early lung cancer among military personnel (DECAMP) consortium: Study protocols, BMC Pulm Med 19 (2019) 1–9. https://doi.org/10.1186/S12890-019-0825-7.

[34] A. McWilliams, M.C. Tammemagi, J.R. Mayo, H. Roberts, G. Liu, K. Soghrati, K. Yasufuku, S. Martel, F. Laberge, M. Gingras, S. Atkar-Khattra, C.D. Berg, K. Evans, R. Finley, J. Yee, J. English, P. Nasute, J. Goffin, S. Puksa, L. Stewart, S. Tsai, M.R. Johnston, D. Manos, G. Nicholas, G.D. Goss, J.M. Seely, K. Amjadi, A. Tremblay, P. Burrowes, P. MacEachern, R. Bhatia, M.-S. Tsao, S. Lam, Probability of Cancer in Pulmonary Nodules Detected on First Screening CT (BROCK), Http://Dx.Doi.Org/10.1056/NEJMoa1214726 369 (2013) 910–919. https://doi.org/10.1056/NEJMOA1214726.

[35] M.N. Kammer, A.K. Kussrow, R.L. Webster, H. Chen, M. Hoeksema, R. Christenson, P.P. Massion, D.J. Bornhop, Compensated Interferometry Measures of CYFRA 21-1 Improve Diagnosis of Lung Cancer, ACS Comb Sci (2019). https://doi.org/10.1021/ACSCOMBSCI.9B00022.

**Table 1. Cohort Inclusion/exclusion criteria**

| | Inclusion/exclusion criteria |
|---|---|
| **NLST-test** | Subjects correspond to the Ardila et al.[18] test set. These subjects were not seen by any of the predictive models in this study. Lung cancer events were the biopsy-confirmed lung cancers reported by the NLST. Subjects without a confirmed outcome were excluded from this study. |
| **NLST-test-nodule** | Subset of the NLST-test cohort in which included patients had at least one positive nodule finding from their CTs as defined in the NLST (>= 4 mm). |
| **NLST-dev** | All subjects enrolled in the CT arm of the NLST and not part of NLST-test. Those without available imaging or without a confirmed outcome were excluded. |
| **LS-A** | Subjects meeting the American Cancer Society criteria for lung screening and were enrolled in the lung screening program at medical center A from 2015 to 2018. Subjects receive longitudinal follow-up after a positive screen and lung cancer events were confirmed via biopsy reports. Nodule characteristics are missing because radiology reports were not available. |
| **LI-A** | Subjects from medical center A who acquired three chest CTs within five years between 2012 and 2019. These subjects were identified through International Classification of Diseases (ICD) codes to have a pulmonary nodule and no cancer before the nodule. We defined lung cancer outcomes through ICD codes representing any malignancy found in the bronchus or lung parenchyma, including metastases from other sites [32]. Nodule characteristics are missing because radiology reports were not available. |
| **Consortium-A** | Prospectively enrolled patients from medical center A and its associated VA between 2003 and 2017. Cohorts prefixed with consortium- meet the following inclusion criteria. Subjects must be aged 18-80 and were detected incidentally to have a pulmonary nodule with a diameter between 6-30mm. Subjects consented at initial nodule detection and serum and CT scan were acquired at that time. Longitudinal imaging and biopsy-confirmed diagnosis for malignant nodules were collected during a 2-year period following initial nodule detection. |
| **Consortium-B** | Prospectively enrolled patients from medical center B according to consortium inclusion criteria. Longitudinal imaging after initial nodule detection was not available. |
| **Consortium-DECAMP** | Prospectively enrolled patients from 12 clinical centers associated with the DECAMP [33] study protocol. Of note, cases and controls are matched on nodule size. |
| **Consortium-C** | Prospectively enrolled patients from medical center C according to MCL inclusion criteria. Longitudinal imaging after initial nodule detection was not available. |
| **BRONCH** | Prospectively collected cohort of subjects who underwent a bronchoscopic lung biopsy for a pulmonary nodule (defined as a lesion < 3cm) at medical center A between the years of 2017-19. The subsequent biopsy report from the bronchoscopy was used to determine benign vs. malignant status of the nodule. |

**Table 2. Cohort Characteristics**

| Cohort | NLST-dev | NLST-test | NLST-test-nodule | LS-A | LI-A | Consortium-A | Consortium-B | Consortium-DECAMP | Consortium-C | BRONCH |
|---|---|---|---|---|---|---|---|---|---|---|
| **Program Type** | Screening | Screening | Screening | Screening | Screening, Incidental | Incidental | Incidental | Incidental | Incidental | Bronchoscopy |
| **Institution** | Multi-institute | Multi-institute | Multi-institute | Medical Center A | Medical Center A | Medical Center A, VA A | Medical Center B | Multi-institute | Medical Center C | Medical Center A |
| **Program Period** | 2002-09 | 2002-09 | 2002-09 | 2015-18 | 2012-21 | 2003-17 | 2006-15 | 2013-17 | 2010-18 | 2017-19 |
| **No. of subjects** | 5436 | 898 | 896 | 882 | 219 | 364 | 117 | 131 | 115 | 373 |
| with lung cancer | 901 (17%) | 149 (17%) | 147 (16%) | 24 (3.0%) | 37 (17%) | 238 (65%) | 48 (41%) | 64 (49%) | 57 (50%) | 230 (62%) |
| **No. of scans** | 14748 | 2523 | 2440 | 1483 | 515 | 760 | 117 | 241 | 115 | 387 |
| with lung cancer | 1866 (13%) | 313 (12%) | 298 (12%) | 51 (3.4%) | 50 (10%) | 517 (68%) | 48 (41%) | 100 (41%) | 57 (50%) | 240 (62%) |
| **Slice thickness (mm)** | 2.1±0.65 | 2.1±0.42 | 2.1±0.42 | 0.81±0.21 | 0.77±0.61 | 1.8±1.1 | 2.2±0.69 | 1.7±0.91 | 1.4±0.79 | 0.9±0.38 |
| **Age** | 62 ± 5.2 | 62 ± 5.2 | 62 ± 5.2 | 65 ± 5.8 | 59 ± 13 | 69 ± 11 | 68 ± 8.5 | 68 ± 7.9 | 66 ± 8.3 | 64 ± 12 |
| **Sex (male)** | 3270 (60%) | 546 (61%) | 546 (61%) | 483 (55%) | 109 (50%) | 165 (45%) | 49 (42%) | 30 (23%) | 31 (27%) | 168 (45%) |
| **BMI** | 28 ± 4.8 | 28 ± 4.9 | 28 ± 5.0 | 28.4 ± 6.0 | 27 ± 7.4 | 28 ± 6.5 | 28 ± 4.9 | 26 ± 5.4 | 29 ± 6.2 | 28 ± 6.8 |
| **Personal cancer history** | 256 (4.7%) | 43 (4.8%) | 43 (4.8%) | 135 (15%) | N/A | 129 (35%) | 3 (2.6%) | 65 (50%) | 13 (11%) | 194 (52%) |
| **Family lung cancer history** | 1194 (22%) | 179 (20%) | 177 (20%) | 149 (17%) | N/A | 41 (11%) | 0 | 0 | 10 (8.7%) | 88 (24%) |
| **Smoking status** | | | | | N/A | | | | | |
| Never | 0 | 0 | 0 | 0 | | 33 (9%) | 0 | 11 (8.4%) | 22 (19%) | 90 (24%) |
| Former | 2781 (51%) | 469 (52%) | 468 (52%) | 357 (40%) | | 195 (54%) | 76 (65%) | 67 (51%) | 54 (47%) | 214 (57%) |
| Current | 2655 (49%) | 429 (48%) | 428 (48%) | 525 (60%) | | 121 (33%) | 41 (35%) | 53 (40%) | 39 (34%) | 69 (18%) |
| **Smoking pack-years** | 56 ± 25 | 59 ± 28 | 59 ± 28 | 48 ± 21 | N/A | 47 ± 33 | 48 ± 23 | 50 ± 25 | 50 ± 33 | 29 ± 30 |
| **Nodule size (mm)** | 8.0 ± 6.2 | 7.9 ± 5.9 | 7.9 ± 5.9 | N/A | N/A | 19 ± 13 | 16 ± 8.9 | 15 ± 7.0 | 18 ± 15 | 2.2 ± 1.3 |
| **Nodule count** | 1.2 ± 1.3 | 1.3 ± 1.2 | 1.3 ± 1.2 | N/A | N/A | 1.0 ± 0.0 | 1.0 ± 0.0 | 1.0 ± 0.0 | 1.0 ± 0.0 | 1.0 ± 0.0 |
| **Nodule attenuation** | | | | N/A | N/A | | | | | |
| Solid | 7494 (51%) | 1351 (54%) | 1351 (55%) | | | 725 (95%) | 99 (85%) | 241 (100%) | 115 (100%) | 325 (84%) |
| Part-solid | 507 (3.4%) | 69 (2.7%) | 69 (2.8%) | | | 21 (2.8%) | 18 (15%) | 0 | 0 | 51 (13%) |
| Non-solid or GGO | 1439 (10%) | 241 (9.6%) | 241 (9.9%) | | | 14 (1.8%) | 0 | 0 | 0 | 11 (2.8%) |
| **Nodule spiculation (present)** | 997 (6.7%) | 200 (7.9%) | 200 (8.2%) | N/A | N/A | 229 (30%) | 15 (13%) | 126 (52%) | 30 (26%) | 173 (45%) |
| **Nodule location** | | | | N/A | N/A | | | | | |
| Upper lobe | 5554 (38%) | 996 (39%) | 996 (41%) | | | 447 (59%) | 63 (54%) | 143 (59%) | 71 (62%) | 203 (52%) |
| Lower lobe | 4391 (30%) | 777 (31%) | 777 (32%) | | | 313 (41%) | 54 (46%) | 98 (41%) | 44 (38%) | 184 (48%) |

**Table 3. Predictive Model Characteristics**

| Model | Year Published | Input | Training Distribution | Cancer Prevalence | Outcome Criteria | Approach |
|---|---|---|---|---|---|---|
| Mayo [8] | 1997 | Age, PH, SS, NSpic, NUL, NSize† | Mayo Clinic (n=419) | 23% | 2-year LC risk proven via tissue biopsy or no findings in follow up | Logistic regression |
| Brock [9] | 2013 | Age, Sex, FH, Emp, Nsize, Nspic NUL, Ncount, Ntype‡ | PanCan [34] (n=1871) | 5.5% | 2-year LC risk proven via tissue biopsy or no findings in follow up | Logistic regression |
| Liao [14] | 2017 | Single chest CT | NLST-dev (n=5436) | 17% | 1-year LC risk proven via tissue biopsy or no findings in follow up | ResNet, nodule detection and ROI-based prediction |
| Sybil [15] | 2023 | Single chest CT | NLST-dev (n=12672) | 17% | Up to 6-year LC risk proven via tissue biopsy or no findings in follow up | ResNet, global chest features and guided attention |
| DLSTM [16] | 2020 | Longitudinal chest CT | NLST-dev (n=5436) | 17% | 6-year LC risk proven via tissue biopsy or no findings in follow up | LSTM network, ROI-based prediction, encodes time interval between scans |
| TdViT [17] | 2023 | Longitudinal chest CT | NLST-dev (n=5436) | 17% | 6-year LC risk proven via tissue biopsy or no findings in follow up | Transformer network, ROI-based prediction, encodes time interval between scans |
| DeepLungScreening [20] | 2021 | Single chest CT, Age, Education, BMI, PH, FH, SS, Quit, PYR | NLST-dev (n=5436) | 17% | 2-year LC risk proven via tissue biopsy or no findings in follow up | ResNet, ROI-based prediction, late fusion of imaging and clinical features |
| DeepLungIPN [19] | 2021 | Single chest CT, Age, BMI, PH, SS, PYR, Nsize, NSpic, NUL, Serum biomarker¶ | Consortium cross validation§ (n=1232) | 59% | 2-year LC risk proven via tissue biopsy or no findings in follow up | DeepLungScreening, serum biomarker |

\* PH: Personal history of any cancer, FH: Family history of lung cancer, SS: smoking status (former v.s. current smoker), SI: smoking intensity (average number of cigarettes smoked a day), SD: smoking duration, Quit: years since the person quit smoking, PYR: pack-years of smoking

†NSpic: Nodule spiculation present or absent, NUL: Nodule in the upper lobes, Nodule Size: largest diameter in mm

‡Emp: Presence of emphysema, Ncount: Number of nodules, Ntype: nodule type, categorized as (1) nonsolid or with ground-glass opacity, (2) part-solid, and (3) solid.

¶Serum concentration of hs-CYFRA 21-1 (natural log of ng/ml) [35]

§Combination of Consortium-A, Consortium-DECAMP, Consortium-C

Table 4: Classification of n-year lung cancer risk using methods across selected cohorts, reported as bootstrap mean AUC [95% CI]

| | | NLST-test (n=898) | NLST-test-nodules (n=896) | LS-A (n=882) | LI-A (n=219) | Consortium-A (n=364) | Consortium-B (n=117) | Consortium-DECAMP (n=131) | Consortium-C (n=115) | BRONCH (n=373) | Average Rank (range) n=‰ |
|---|---|---|---|---|---|---|---|---|---|---|---|
| | Years to outcome | 2 year risk | 2 year risk | 2 year risk | 3 year risk | 2 year risk | 2 year risk | 2 year risk | 2 year risk | 1 year risk | |
| **Input** | **Method** | | | | | | | | | | |
| Clinical Variables | **Mayo** | ‡ | 0.804 [0.798, 0.809] | ‡ | ‡ | 0.706 [0.704, 0.708] | 0.864 [0.862, 0.867] | 0.568 [0.565, 0.571] | 0.716 [0.712, 0.719] | 0.621 [0.615, 0.628] | 3.5 (7, 1) n=6 |
| | **Brock** | ‡ | 0.789 [0.782, 0.796] | ‡ | ‡ | 0.716 [0.714, 0.718] | 0.885 [0.883, 0.886] | 0.662 [0.659, 0.666] | 0.713 [0.710, 0.716] | 0.497 [0.494, 0.499] | 3.2 (5, 2) n=6 |
| Single Chest CT | **Liao et al.** | 0.751 [0.747, 0.756] | 0.755 [0.750, 0.759] | 0.723 [0.712, 0.734] | 0.644 [0.635, 0.653] | 0.662 [0.660, 0.664] | 0.779 [0.776, 0.782] | 0.706 [0.703, 0.709] | 0.660 [0.656, 0.663] | 0.621 [0.614, 0.628] | 3.9 (6, 1) n=9 |
| | **Sybil** | **0.881 [0.877, 0.885]*** | **0.879 [0.872, 0.885]*** | 0.779 [0.768, 0.789] | 0.670 [0.661, 0.679] | 0.700 [0.694, 0.706] | 0.889 [0.884, 0.895] | 0.606 [0.597, 0.616] | 0.764 [0.756, 0.772] | **0.623 [0.618, 0.629]** | 2.6 (6, 1) n=9 |
| Multiple Chest CTs | **DLSTM** | 0.738 [0.734, 0.743] | 0.727 [0.721, 0.731] | ¶ | 0.711 [0.702, 0.720] | 0.743 [0.741, 0.745] | § | 0.778 [0.774, 0.781] | § | § | 3.6 (6, 2) n=5 |
| | **TDViT** | 0.797 [0.793, 0.802] | 0.790 [0.785, 0.794] | ¶ | **0.773 [0.764, 0.781]*** | 0.753 [0.750, 0.755] | § | **0.823 [0.820, 0.825]*** | § | § | 1.8 (3, 1) n=5 |
| Single Chest CT and clinical variables | **DLS** | 0.783 [0.778, 0.788] | 0.776 [0.771, 0.782] | **0.810 [0.799, 0.820]*** | † | † | † | † | † | † | 2.7 (4, 1) n=3 |
| | **DLI** | † | † | † | † | **0.856 [0.854, 0.858]*** | **0.936 [0.935, 0.938]*** | 0.742 [0.739, 0.745] | **0.851 [0.849, 0.854]** | † | 1.5 (3, 1) n=4 |

* result was significantly different compared to each other method in the column for p<0.01
‡ nodule characteristics unavailable (missing >10% of nodule size, attenuation, count, spiculation, or lobe location)
¶ prohibitive class imbalance (only 6/23 lung cancer cases have more than one scan)
† Missing demographic, smoking history, COPD, or CYFRA covariates
§ No longitudinal imaging
‰ Number of cohort evaluations performed with this model

**Table 5: Classification of lung cancer risk by subtype.**

|  | NLST-test-nodules (n malignant=147, n benign=749) | | Consortium-A (n malignant=238, n benign=126) | |
| --- | --- | --- | --- | --- |
|  | SCLC (n=18) | NSCLC (n=120) | SCLC (n=39) | NSCLC (n=194) |
| **Mayo** | ¶ (n=5) | 0.810 [0.808, 0.812] | **0.774 [0.772, 0.776]*** | 0.683 [0.681, 0.685] |
| **Brock** | ¶ (n=5) | 0.792 [0.790, 0.794] | **0.794 [0.792, 0.797]*** | 0.688 [0.687, 0.690] |
| **Liao et al.** | 0.683 [0.680, 0.687] | **0.770 [0.768, 0.771]*** | 0.617 [0.614, 0.620] | **0.688 [0.686, 0.690]*** |
| **Sybil** | 0.728 [0.723, 0.733] | **0.899 [0.897, 0.900]*** | 0.701 [0.698, 0.703] | 0.701 [0.699, 0.702] |
| **DLSTM** | 0.663 [0.658, 0.668] | **0.808 [0.806, 0.809]*** | 0.730 [0.726, 0.735] | **0.754 [0.751, 0.757]*** |
| **TDViT** | 0.707 [0.702, 0.711] | **0.771 [0.769, 0.773]*** | 0.667 [0.661, 0.673] | **0.760 [0.757, 0.763]*** |
| **DLS** | 0.659 [0.654, 0.664] | **0.792 [0.791, 0.794]*** | † | † |
| **DLI** | † | † | **0.904 [0.901, 0.907]*** | 0.853 [0.851, 0.855] |

\* result was significantly different compared to each other method in the column for p<0.01
¶ prohibitive class imbalance
† Missing demographic, smoking history, COPD, or CYFRA covariates

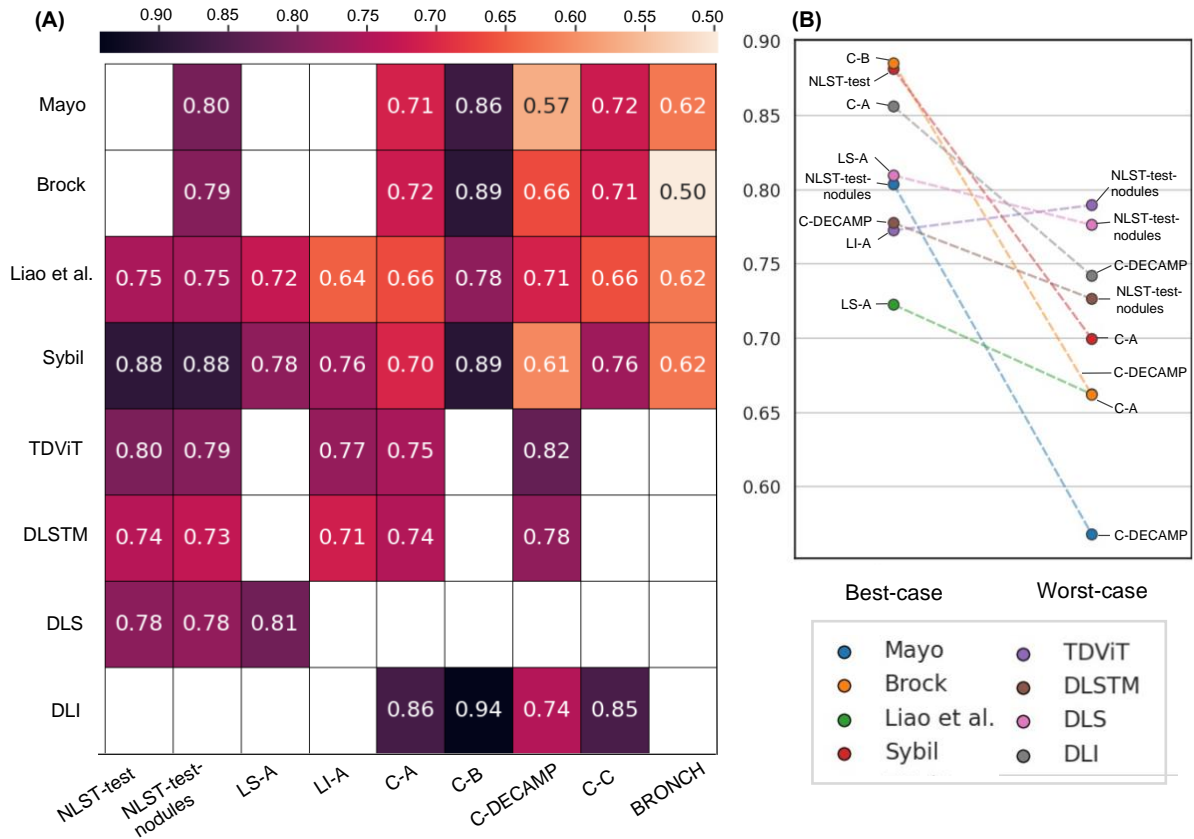

**Figure 1. A.** Mean AUC for all methods applied on all study cohorts. Almost all methods demonstrate a high degree of variance in performance across cohorts within most methods, which demonstrates the importance of contextualizing a models performance by comparing it with the performance of baseline models. 95% CIs visualized in Supp. Fig 2. **B.** Best- and worst-case performance for 8 predictive models reveals robust performance of longitudinal and multimodal AI methods (i.e. TDViT, DLSTM, DLS, DLI) compared to other models. A model's worst case performance is defined as its lowest ranked performance across all cohorts except BRONCH.

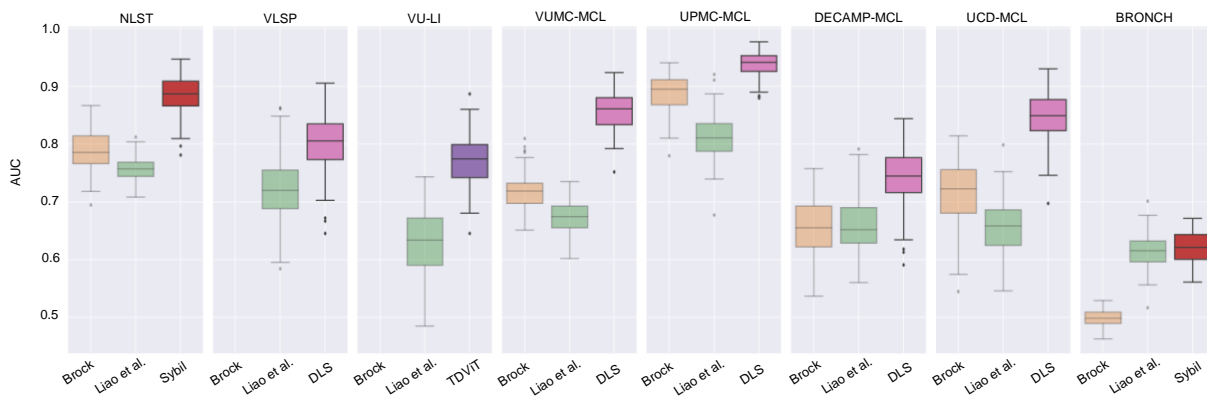

**Figure 2.** AUC of 100 bootstrapped samples from applying three selected methods across all study cohorts. Brock and Liao et al. are selected as baselines to compare with the method achieving the highest classification performance in the corresponding cohort. The best performing method differs across cohorts. Among baselines and the best performers, bootstrapped AUC distributions demonstrate high variance across cohorts. DLS seems to perform the best in cohorts with incidental nodules. Unsurprisingly, TDViT excels in the longitudinal imaging cohort (LI-A).

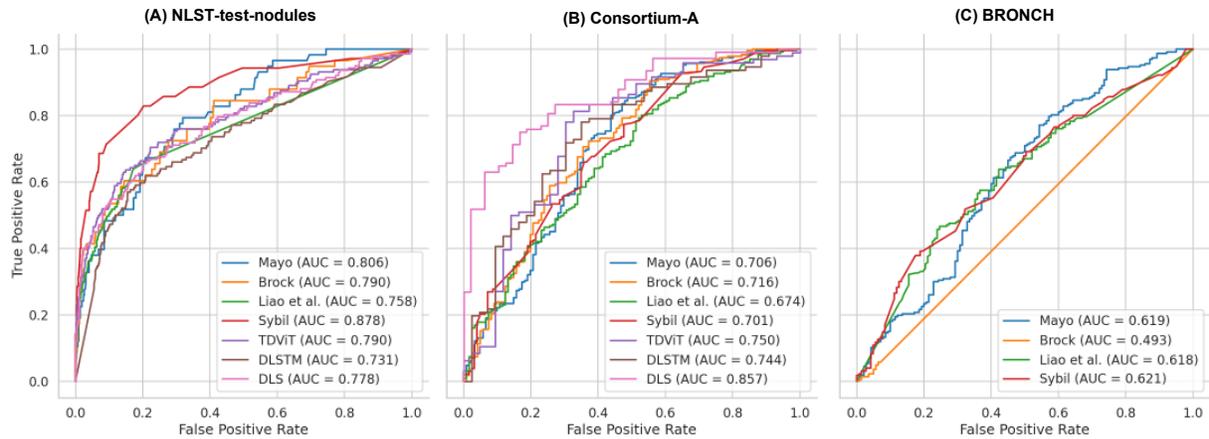

**Figure 3.** The receiver operating characteristic (ROC) curves demonstrate a failure to generalize across four select cohorts. Top performers in lung screening cohorts (**A**) are different than the top performers in cohorts with incidentally-detected nodules (**B**), and vice-versa. (**C**) All evaluated models performed poorly on a retrospective cohort of subjects selected to undergo diagnostic bronchoscopic biopsy for a pulmonary nodule. ROC curves for remaining evaluation cohorts are provided in Supp. Fig 1.